\documentclass[twocolumn,prl,superscriptaddress,showpacs]{revtex4}
\usepackage{epsf,graphicx,amssymb}

\begin{document}
\draft
\title{Generation of magnetic field by dynamo action in a turbulent flow of liquid sodium}

\author{R. Monchaux}
\affiliation{Service de Physique de l'Etat Condens\'e, Direction des Sciences de la Mati\`ere, CEA-Saclay, CNRS URA 2464, 91191 Gif-sur-Yvette cedex, France}
\author{M. Berhanu}
\affiliation{Laboratoire de Physique Statistique de l'Ecole Normale
Sup\'erieure, CNRS UMR 8550, 24 Rue Lhomond, 75231 Paris Cedex 05, France}
\author{M. Bourgoin}
\altaffiliation[Present address : ]{LEGI, CNRS UMR 5519, BP53, 38041 Grenoble, France}
\author{M. Moulin}
\author{Ph. Odier}
\author{J.-F. Pinton}
\author{R. Volk}
\affiliation{Laboratoire de Physique de l'Ecole Normale
Sup\'erieure de Lyon, CNRS UMR 5672, 46 all\'ee d'Italie, 69364 Lyon 
Cedex 07, France}
\author{S. Fauve}
\affiliation{Laboratoire de Physique Statistique de l'Ecole Normale
Sup\'erieure, CNRS UMR 8550, 24 Rue Lhomond, 75231 Paris Cedex 05, France}
\author{N. Mordant}
\author{F. P\'etr\'elis}
\affiliation{Laboratoire de Physique Statistique de l'Ecole Normale
Sup\'erieure, CNRS UMR 8550, 24 Rue Lhomond, 75231 Paris Cedex 05, France}
\author{A. Chiffaudel}
\author{F. Daviaud}
\author{B. Dubrulle}
\author{C. Gasquet}
\author{L. Mari\'e}
\altaffiliation[Present address : ]{IFREMER, Laboratoire de Physique des Oc\'eans, CNRS UMR 5519, BP70, 29280 Plouzane, France}
\author{F. Ravelet}
\altaffiliation[Present address : ]{Laboratory for Aero and Hydrodynamics, TU-Delft, The Netherlands}
\affiliation{Service de Physique de l'Etat Condens\'e, Direction des Sciences de la Mati\`ere, CEA-Saclay, CNRS URA 2464, 91191 Gif-sur-Yvette cedex, France}

\date{\today}

\begin{abstract}
We report the observation of dynamo action in the VKS experiment, i.e., the generation of magnetic field by a strongly turbulent swirling flow of liquid sodium. Both mean and fluctuating parts of the field are studied. The dynamo threshold corresponds to a magnetic Reynolds number $R_m \sim 30$. A mean magnetic field of order $40$ G is observed $30\, \%$ above threshold at the flow lateral boundary. The $rms$ fluctuations are larger than the corresponding mean value for two of the components. The scaling of the mean square magnetic field is compared to a prediction previously made for high Reynolds number flows.

\end{abstract}
\maketitle
The generation of electricity from mechanical work has been one of the main achievements of physics by the end of the $\rm{XIX^{th}}$ century.  In 1919, Larmor proposed that a similar process can generate the magnetic field of the sun from the motion of an electrically conducting fluid. However, fluid dynamos are more complex than industrial ones and it is not easy to find laminar flow configurations that generate magnetic fields \cite{mof78}. Two simple but clever examples have been found in the seventies \cite{robpom} and have led more recently to successful experiments \cite{exp01}. These experiments have shown that the observed thresholds are in good agreement with theoretical predictions \cite{laminarpredict} made by considering only the mean flow, whereas the saturation level of the magnetic field cannot be described with a laminar flow model (without using an ad-hoc turbulent viscosity) \cite{petrelis01}. These observations have raised many questions: what happens for flows without geometrical constraints such that fluctuations are of the same order of magnitude as the mean flow? Is the dynamo threshold strongly increased due to the lack of coherence of the driving flow \cite{schekochihin04,laval06} or does the prediction made as if the mean flow were acting alone still give a reasonable order of magnitude \cite{ponty05}? What is the nature of the dynamo bifurcation in the presence of large velocity fluctuations? All these questions, and others motivated by geophysical or astrophysical dynamos \cite{gafd}, have led several teams to try to generate dynamos in flows with a high level of turbulence \cite{nondynamos,vks}.  
We present in this letter our first experimental observation of the generation of a magnetic field in a von K\'arm\'an swirling flow of liquid sodium (VKS) for which velocity fluctuations and the mean flow have comparable  kinetic energy and we discuss some of the above issues.\\
\begin{figure}[h]
\centerline{
\epsfysize=60mm
\epsffile{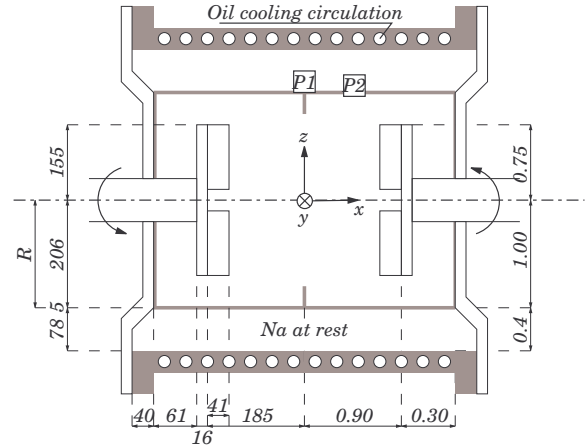} }
\caption{Sketch of the experimental set-up. The inner and outer cylinders are made of copper (in gray). The dimension are given in millimeter (left) and normalized by  R (right). The 3D Hall probe is located either at point P1 in the mid-plane or P2. In both cases, the probe is nearly flush with the inner shell.}
\label{Fig1}
\end{figure}
The experimental set-up (see Fig. \ref{Fig1}) is similar to the previous VKS experiments \cite{vks}, but involves three  modifications that will be described below. 
The flow is generated by rotating two disks of radius $154.5$ mm, $371$ mm apart in a cylindrical vessel, $2R = 412$ mm in inner diameter and $524$ mm in length. The disks are fitted with 8 curved blades of height $h=41.2$ mm. These impellers are driven at a rotation frequency up to $\Omega/2\pi=26$ Hz by $300$ kW available motor power. An oil circulation in the outer copper cylinder maintains a regulated temperature in the range $110$-$160\, ^{o}$C. The mean flow has the following characteristics: the fluid is ejected radially outward by the disks;  this drives an axial flow toward the disks along their axis and a recirculation in the opposite direction along the cylinder lateral boundary. In addition, in the case of counter-rotating disks studied here, the presence of a strong axial shear of azimuthal velocity in the mid-plane between the impellers generates a high level of turbulent fluctuations \cite{marie,note}.
 The kinetic Reynolds number is $Re= K R^2 \Omega/\nu$, where $\nu$ is the kinematic viscosity and $K = 0.6$ is a coefficient that measures the efficiency of the impellers \cite{ravelet05}. $Re$ can be increased up to  $5\, 10^6$: the corresponding magnetic Reynolds number is, $R_m = K \mu_0 \sigma R^2 \Omega \approx 49$ (at $120\, ^{o}$C), where $\mu_0$ is the magnetic permeability of vacuum and $\sigma$ is the electrical conductivity of sodium. 

A first modification with respect to earlier VKS experiments consists of  surrounding the flow by sodium at rest in another concentric cylindrical vessel, $578$ mm in inner diameter. This has been shown to decrease the dynamo threshold in kinematic computations based on the mean flow velocity \cite{ravelet05}. The total volume of liquid sodium is $150$ l. 
A second geometrical modification consists of attaching an annulus of inner diameter $175$ mm and thickness $5$ mm along the inner cylinder in the mid-plane between the disks. Water experiments have shown that its effect on the mean flow is to make the shear layer sharper around the mid-plane. In addition, it reduces low frequency turbulent fluctuations, thus the large scale flow time-averages faster toward the mean flow. However, $rms$ velocity fluctuations are almost unchanged (of order $40 - 50 \%$), thus the flow remains strongly turbulent \cite{ravelet05b}. It is expected that reducing the transverse motion of the shear layer decreases the dynamo threshold for the following reasons: (i) magnetic induction due to an externally applied field on a gallium flow strongly varies because of the large scale flow excursions away from the time averaged flow \cite{volk06}, (ii) the addition of large scale noise to the Taylor-Green mean flow  increases its dynamo threshold \cite{laval06}, (iii) fluctuating motion of eddies increase the dynamo threshold of the Roberts flow \cite{phase06}.

\begin{figure}[h]
\centerline{
\epsfysize=60mm
\epsffile{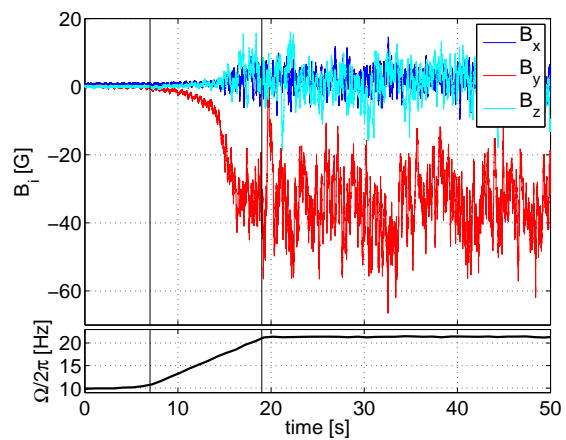} }
\caption{Time recording at $P1$ of the components of the magnetic field when the rotation frequency $\Omega /2\pi$ is increased as displayed by the ramp below ($R_m$ increases from $19$ to $40$).}
\label{Fig2}
\end{figure}

The above configuration does not generate a magnetic field up to the maximum possible rotation frequency of the disks ($\Omega/2\pi=26$ Hz). We thus made a last modification and replaced disks made of stainless steel by similar iron disks. Using boundary conditions with a high permeability in order to change the dynamo threshold has been already proposed \cite{peyresq}. It has been also shown that in the case of a Ponomarenko or G. O. Roberts flows, the addition of an external wall of high permeability can decrease the dynamo threshold \cite{plunian03}. Finally, recent kinematic simulations of the VKS mean flow have shown that different ways of taking into account the sodium behind the disks lead to an increase of the dynamo threshold ranging from $12 \, \%$ to $150 \, \%$ \cite{stefani06}. We thought that using iron disks could screen magnetic effects in the bulk of the flow from the region behind the disks, although the actual behavior may be more complex.  
This last modification generates a dynamo above $R_m \simeq 30$. The three components of the field $\vec{B}$ are measured with a 3D Hall probe, located either in the mid-plane or $109$ mm away from it ($P1$ or $P2$ in Fig. \ref{Fig1}). In both cases, the probe is nearly flush with the inner shell, thus $\vec{B}$ is measured at the boundary of the turbulent flow. Fig. \ref{Fig2} shows the time recording of the three components of $\vec{B}$ when $R_m$ is increased from $19$ to $40$. The largest component, $B_y$, is tangent to the cylinder at the measurement location. It  increases from a mean value comparable to the Earth magnetic field to roughly $40$ G. The mean values of the other components $B_x$ and $B_z$ also increase (not visible on the figure because of fluctuations). Both signs of the components have been observed in different runs, depending on the sign of the residual magnetization of the disks. All components display strong fluctuations as could be expected in flows with Reynolds numbers larger than $10^{6}$.

\begin{figure}[h]
\centerline{
\includegraphics[width=.45\textwidth,height=.27\textheight]{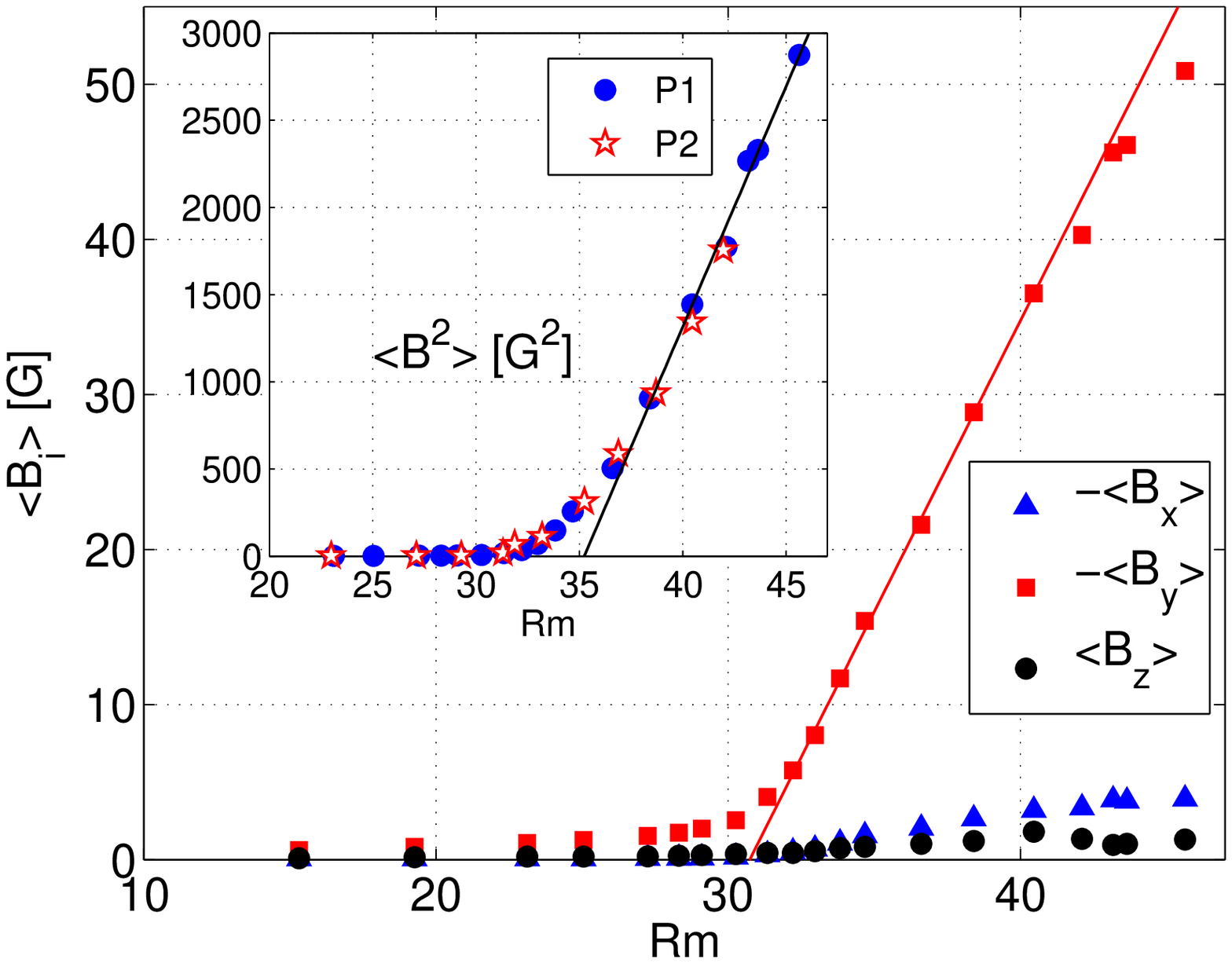}}
\centerline{
\includegraphics[width=.41\textwidth,height=.25\textheight]{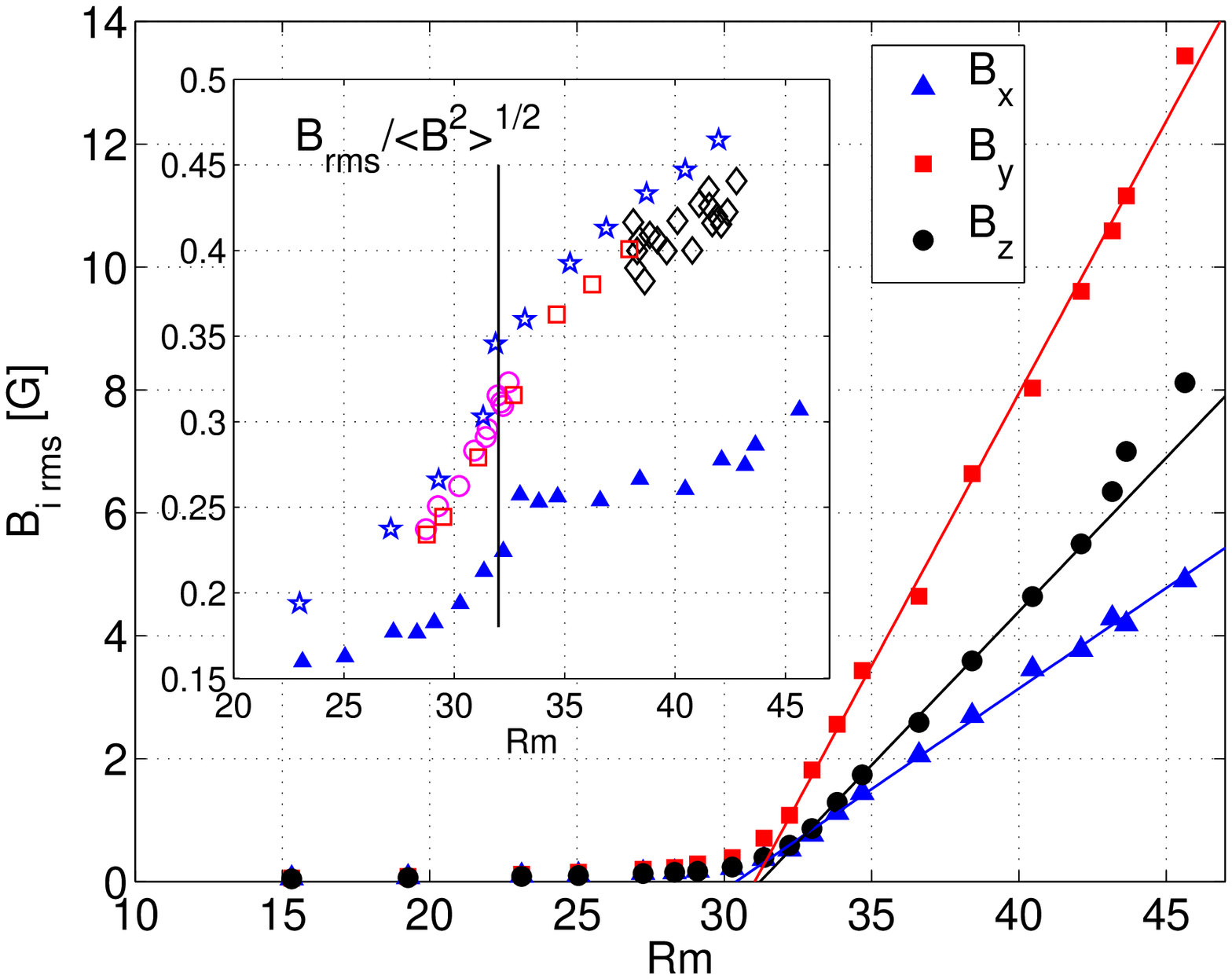}}
\caption{a) Mean values of the three components of the magnetic field recorded at $P1$ versus $R_m$ ($T = 120\, ^{o}$C): ($\blacktriangle$) $-\langle B_x \rangle$, ($\blacksquare$) $-\langle B_y \rangle$, ($\bullet$) $\langle B_z \rangle$. The inset shows the time average of the square of the total magnetic field as a function of $R_m$, measured at $P1$ ($\bullet$), or at P2 ($\star$) after being divided by $1.8$.
b) Standard deviation of the fluctuations of each components of the magnetic field recorded at $P1$ versus $R_m$.
The inset shows $B_{rms}/\langle B^2 \rangle^{1/2}$. Measurements done at $P1$:  ($\blacktriangle$) $T = 120\, ^{o}$C,  frequency increased up to $22$ Hz; Measurements done at $P2$: ($\star$) $T = 120\, ^{o}$C,  frequency decreased from $22$ to $16.5$ Hz,   ($\square$) $T = 156\, ^{o}$C, frequency increased up to $22$ Hz,  ($\circ$) $\Omega /2\pi = 16.5$ Hz, $T$ varied from $154$ to $116\, ^{o}$C, ($\lozenge$) $\Omega /2\pi = 22$ Hz, $T$ varied from $119$ to $156\, ^{o}$C. The vertical line corresponds to $R_m = 32$.}
\label{Fig3}
\end{figure}

Fig. \ref{Fig3}a shows the mean values of the components $\langle B_i \rangle$ of the magnetic field and Fig. \ref{Fig3}b their fluctuations $B_{i\, rms}$ versus $R_m$. The fluctuations are all in the same range ($3$ G to $8$ G, at  $30 \, \%$ above threshold) although the corresponding mean values are very different. The time average of the square of the total magnetic field, $\langle \vec{B}^2 \rangle$, is displayed in the inset of Fig. \ref{Fig3}a. No hysteresis is observed. Linear fits of $\langle B_y \rangle$ or $B_{i\, rms}$ displayed in Fig. \ref{Fig3} define a critical magnetic Reynolds number $R_m^{c} \sim 31$ whereas the linear fit of $\langle \vec{B}^2 \rangle$ gives a larger value $R_m^{0} \sim 35$. The latter is the one that should be considered in the case of a supercritical pitchfork bifurcation. The rounding observed close to threshold could then be ascribed to the imperfection due to the ambient magnetic field (Earth field, residual magnetization of the disks and other magnetic perturbations of the set-up). The actual behavior may be more complex because this bifurcation takes place on a strongly turbulent flow, a situation for which no rigorous theory exists. The inset of  Fig. \ref{Fig3}b shows that the variance $B_{rms}^2 = \langle (\vec{B}  - \langle \vec{B} \rangle)^2 \rangle$ is not proportional 
to $\langle B^2 \rangle$. Below the dynamo threshold, the effect of induction due to the ambient magnetic field is observed. $B_{rms}/\langle B^2 \rangle^{1/2}$ first behaves linearly at low $R_m$, but then increases faster as $R_m$ becomes closer to the bifurcation threshold.   We thus show that this seems to be a good quantity to look at as a precursor of a dynamo regime. In addition, we observe that it displays a discontinuity in slope in the vicinity of  $R_m^{c}$ in an analogous way of some response functions at phase transitions or bifurcations in the presence of noise. Note however that the shape of the curves depends on the measurement point and they cannot be superimposed with a scaling factor as done for $\langle B^2 \rangle$ versus $R_m$ in the inset of Fig. \ref{Fig3}a.

The above results are characteristic of bifurcations in the presence of noise. As shown in much simpler experiments, different choices of an order parameter (mean value of the amplitude of the unstable mode or its higher moments, its most probable value, etc) can lead to qualitatively different bifurcation diagrams \cite{berthet}.  This illustrates the ambiguity in the definition of the order parameter for bifurcations in the presence of fluctuations or noise. In the present experiment, fluctuations enter both multiplicatively, because of the turbulent velocity,  and additively, due to the interaction of the velocity field with the ambient magnetic field.  Finally, we note that both $R_m^{c}$ and $R_m^{0}$ are smaller than the thresholds computed with kinematic dynamo codes taking into account only the mean flow, that are in the range $R_m^{c} = 43$ to $150$ depending on different boundary conditions on the disks and on configurations of the flow behind them \cite{ravelet05,stefani06}.

The probability density functions (PDF) of the fluctuations of the three components of the induced magnetic field (not displayed) are roughly gaussian.  The PDFs of fluctuations below threshold, i.e., due to the induction resulting from the ambient magnetic field, are similar to the ones observed in the self-generating regime. We do not observe any non gaussian behavior close to threshold which would result from an on-off intermittency mechanism \cite{sweet01}. Possible reasons are the low level of small frequency velocity fluctuations \cite{aumaitre05} or the imperfection of the bifurcation that results from the ambient magnetic field \cite{petrelis06}.

\begin{figure}[h]
\centerline{
\epsfysize=65mm
\epsffile{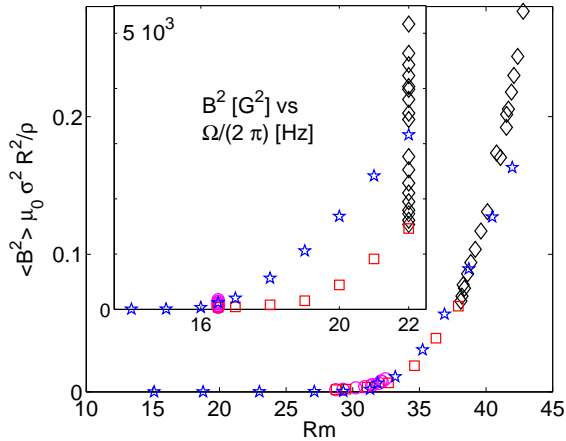} }
\caption{The dimensionless quantity, $\langle B^2 \rangle  \mu_0 (\sigma R)^2 / \rho$ is displayed as a function of $R_m$ for different working temperatures and frequencies (measurements done at $P2$ and identical symbols as in the inset of Fig. \ref{Fig3}b). The inset shows the same data in dimensional form $B^2$ versus rotation frequency for different temperatures.} 
\label{Fig4} 
\end{figure}

Fig. \ref{Fig4} displays both the dimensional (see inset) and dimensionless mean square field as a function of $R_m$. $\langle B^2 \rangle$ is made dimensionless using a high Reynolds number scaling \cite{petrelis01}:  $\langle B^2 \rangle \propto \rho/ (\mu_0 (\sigma R)^2) \, (R_m - R_m^{c}) / R_m^{c}$, where $\rho$ is the fluid density. We observe that data obtained at different working temperature are well collapsed by this scaling ($\sigma$ decreases by roughly $15$ $\%$ from $100$ to $160$ $^{o}$C). The low Reynolds number or ``weak field" scaling could also give a reasonable collapse of data obtained on this temperature range but the predicted order of magnitude for $\langle B^2 \rangle$ would be $10^5$ too small.

Dissipated power by Ohmic losses is another important characterization of dynamo action.
Our measurements show that, $30 \, \%$ above threshold, it leads to an excess power consumption of $15-20 \, \%$ with respect to a flow driving power of the order of $100$ kW. 

The effect of iron disks deserves additional discussion. A slight effect of magnetization of iron has been observed: the  dynamo threshold during the first run was about $20 \,\%$ larger than in the next runs for which all the measurements were then perfectly reproducible. However, no effect of remanence that would lead to a
 hysteretic behavior close to the bifurcation threshold has been observed. Demagnetization of pure iron occurring for field amplitudes of the order of the Earth field, i.e., much smaller than the fields generated by the dynamo, the iron disks do not impose any permanent magnetization but mostly change the boundary condition for the magnetic field generated in the bulk of the flow. This changes the dynamo threshold and the near critical behavior for amplitudes below the coercitive field of pure iron. 
It should be also emphasized that the axisymmetry of the set-up cannot lead to Herzenberg-type dynamos \cite{lowes}. In addition, these rotor dynamos display a sharp increase of the field at threshold and their saturation is mostly limited by the available motor power \cite{lowes}. On the contrary, we observe a continuous bifurcation with a saturated magnetic field in good agreement with a scaling law derived for a fluid dynamo. 

The different mechanisms at work, effect of magnetic boundary conditions, effect of mean flow with respect to turbulent fluctuations, etc, will obviously motivate further studies of the VKS dynamo. A preliminary scan of the parameter space has shown that when the disks are rotated at different frequencies, other dynamical dynamo regimes are observed including random inversions of the field polarity. Their detailed description together with experiments on the relative effect of the mean flow and the turbulent fluctuations on these dynamics are currently in progress.

\begin{acknowledgments}
We gratefully aknowledge the assistance of D. Courtiade, J.-B. Luciani, P. Metz, V. Padilla, J.-F. Point and A. Skiara and the participation of J. Burguete to the early stage of VKS experiment. This work is supported by the french institutions: Direction des Sciences de la Mati\`ere and Direction de l'Energie Nucl\'eaire of CEA, Minist\`ere de la Recherche and Centre National de Recherche Scientifique (ANR 05-0268-03, GDR 2060). The experiments have been realized in CEA/Cadarache-DEN/DTN.  
\end{acknowledgments}


\end{document}